\documentclass[]{aa} % for a referee version
\usepackage{graphicx}
\usepackage{txfonts}
\usepackage[utf8]{inputenc}
\usepackage{natbib}
\bibpunct{(}{)}{;}{a}{}{,} % to follow the A&A style
\newcommand{\mygi}{MyGIsFOS}
\newcommand{\Teff}{\ensuremath{T_\mathrm{eff}}}

\newcommand{\loggf}{\ensuremath{\log\,gf}}
\newcommand{\logg}{\ensuremath{\log g}}

\newcommand{\fei}{\ion{Fe}{i}}
\newcommand{\feii}{\ion{Fe}{ii}}
\newcommand{\alphafe}{[$\alpha$/Fe]}
\newcommand{\Vturb}{V$_\mathrm{turb}$}
\newcommand{\feh}{[Fe/H]}
\newcommand{\sgr}{Sgr dSph}
\begin{document} 

\title{A wide angle view of the Sagittarius dwarf spheroidal galaxy}

\subtitle{II. A CEMP-r/s star in the Sagittarius dwarf Spheroidal Galaxy\thanks{Based on data collected with UVES at 8.2 m VLT-UT2 (Kueyen) telescope under ESO programme 083.B-0774. This paper includes data gathered with the 6.5 meter Magellan Telescopes located at Las Campanas Observatory, Chile.}}

   \author{ L. Sbordone\inst{1}
            \and
            C. J. Hansen\inst{2}
            \and
            L. Monaco\inst{3}
            \and
            S. Cristallo\inst{4,5}
            \and
            P. Bonifacio\inst{6}
            \and
            E. Caffau\inst{6}
            \and
            S.Villanova\inst{7}
            \and
            P. Amigo\inst{8}
          }

\institute{ESO - European Southern Observatory, Alonso de Cordova 3107, Vitacura, Santiago, Chile\\
\email{lsbordon@eso.org}
\and
Max Planck Institute for Astronomy, Heidelberg, K\"onigstuhl 17, D-69117 Heidelberg, Germany
\and
Departamento de Ciencias Fisicas, Universidad Andres Bello, Fernandez Concha 700, Las Condes, Santiago, Chile
\and
INAF - Osservatorio Astronomico d'Abruzzo, via Maggini snc, Teramo, Italy
\and
INFN - Sezione di Perugia, via A. Pascoli, Perugia, Italy
\and
GEPI, Observatoire de Paris, Universit\'{e} PSL, CNRS, Place Jules Janssen, 92195 Meudon, France
\and
Departamento de Astronom\'ia, Universidad de Concepci\'on, Casilla 160-C, Concepci\'on, Chile
\and
Facultad de Ingenier\'ia y Ciencias, Universidad Adolfo Ib\'a\~nez, Avenida Padre Hurtado 750, Viña del Mar, 2520000, Chile
}

   \date{Received September 15, 1996; accepted March 16, 1997}

\abstract{
We report on the discovery and chemical abundance analysis of the first CEMP-r/s star detected in the Sagittarius dwarf Spheroidal Galaxy, by means of UVES high resolution spectra. The star, found in the outskirts of \sgr, along the main body major axis, is a moderately metal poor giant (\Teff=4753 K, \logg=1.75, \feh=-1.55), with [C/Fe]=1.13 placing it in the so-called ``high-carbon band'', and strong s-process and r-process enrichment ([Ba/Fe]=1.4, [Eu/Fe]=1.01). Abundances of 29 elements from C to Dy were obtained. The chemical pattern appears to be best fitted by a scenario where an r-process pollution event pre-enriched the material out of which the star was born as secondary in a binary system whose primary evolved through the AGB phase, providing C and s-process enrichment.}

%\begin{document} 
\maketitle
%
%-------------------------------------------------------------------

\section{Introduction}

The Carbon Enhanced Metal-Poor stars (hereafter CEMP) are
objects of low metallicity that show a high abundance of C,
relative to iron. A thorough historical perspective of
carbon enhanced stars is given by \citet{caffau18} and
we refer the reader to that paper, and references therein
for further details. \citet{beers05} introduced a classification
of the CEMP stars based on their abundances of neutron capture
elements, in particular, for what is relevant here, they defined
two classes of stars that appear to be enriched in neutron capture 
elements. They made a distinction whether only the elements
produced in the slow neutron capture process ($s-$process)
or also the nuclei produced
in the rapid neutron capture process ($r-$process)
are enhanced.
Taking into account that most Ba isotopes are produced in the
$s-$process and the vast majority of Eu nuclei are produced in
the $r-$process, \citet{beers05} defined 
CEMP-s stars as stars that satisfy 
[C/Fe]$ > +1.0$, [Ba/Fe] $> +1.0$, and [Ba/Eu]$ > +0.5$
and CEMP-r/s as stars that satisfy 
[C/Fe] $> +1.0$ and $0.0<$[Ba/Eu] $< +0.5$.
\citet{lucatello05} and later \citet{starkenburg14} from repeated
radial velocity measurements of CEMP stars argued that the binary frequency among CEMP-s
stars is compatible with a population composed by 100\% of binary stars. 
\citet{spite12} and \citet{bonifacio18} pointed out that the CEMP stars
have a distinctly bi-modal distribution in carbon abundances A(C)\footnote{We adopt the notation for any element X,
A(X) = log$_{10}$(X/H) +12.}: stars of the low-carbon band (A(C)$\le 7.6$) and
stars of the high-carbon band (A(C)$> 7.6$). The high-carbon band is populated
almost exclusively by stars of the classes CEMP-s and CEMP-r/s. 
Recently \citet{arentsen19} revised the binary frequency among CEMP stars
and estimated a lower frequency than those of \citet{lucatello05} and \citet{starkenburg14},
yet they confirmed that the binary frequency among stars of the high-carbon band (47\%)
is much higher than that among those of the low-carbon band (18\%).
The simplest, and most widely adopted, interpretation of this observational
picture is that the high-carbon band stars are the result of mass transfer in a binary 
system \citep{abate18}. 

For CEMP-s stars it is straightforward to identify the
companion that transferred mass as an Asymptotic Giant Branch (AGB)
star, that can produce both the excess carbon and the neutron capture 
elements \citep[see e.g.][and references therein]{cristallo11,bisterzo12,kappeler11}.
The situation for CEMP-r/s  stars is less clear, since the neutron densities required for 
the $s-$process and the $r-$process differ by at least 10 orders of magnitude \citep[see e.g.][and references therein]{hampel16},
they likely form in different astrophysical sites. 
One model for explaining CEMP-r/s stars is to assume that the cloud out of which
the star was formed had been previously enriched in $r-$process elements \citep[see e.g.][]{bisterzo12},
by a different source, like a neutron-star merger (\citealt{thielemann17}, \citealt{watson19}), a magnetar \citep{siegel19} or a Magneto Hydro-Dynamic (MHD) Supernova \citep{nishimura15}.
Other more contrived scenarios have been proposed and we refer the reader to \citet{hampel16} for
a concise summary of the relevant literature.
Recently, computations have been done to explore nucleosynthesis at  intermediate neutron densities,
in the range $\rm 10^7 - 10^{15} cm^{-3}$ in AGB \citep{hampel16}  and Rapidly Accreting White Dwarfs  \citep[RAWDs,][]{Deni19} to explain the origin of CEMP-r/s stars.
The high end of this neutron density regime was originally investigated by \citet{cr77} 
to explain the production of $^{14}$C and neutrons in red giants, now it
is usually referred to as the $i-$process.

In this paper we report the discovery and analysis of the first CEMP-r/s star
found in the Sgr dSph galaxy. 
The significance of this discovery is to be able to compare
the properties of this particular class of stars to those of the Galactic stars. 
Such a comparison could give information on how the galactic environment
affects, or does not affect, the evolution of these exceptional objects.

\section{Observations}

The star GIU J190734.24-315102.1 (henceforth J1907) was catalogued as part of the Sgr dSph Wide Angle survey \citep[][]{giuffrida10,hansen18,sbordone15}. In \citet{giuffrida10} it was detected as a probable Sgr dSph member from VLT-VIMOS \citep{lefevre03} V and I photometry and low-dispersion spectroscopy. Subsequently, it was followed up with FLAMES-GIRAFFE \citep{pasquini00}, which allowed to firmly establish it as a high-probability member, and to derive an overall metallicity of [Fe/H]$\sim$ -1.5 (Sbordone et al., in preparation). Finally, together with a group of low-metallicity Sgr dSph member stars, it was re-observed with UVES \citep{dekker00} to obtain high-resolution, high-quality spectra for detailed chemical analysis. The other stars re-observed with UVES-slit together with J1907 have been presented in \citet{hansen18}. 

Coordinates, photometry, proper motions and atmospheric parameters for J1907 are listed in Table\,\ref{tableparams}. Coordinates and V, I magnitudes come from \citet{giuffrida10}. J1907 is situated in the easternmost major-axis field described in \citet{giuffrida10}, Sgr4, roughly 3 degrees away from the center of \object{NGC6715} (\object{M54}), which coincides with the center of the \object{Sgr dSph}. While its radial velocity differs from the one of \object{NGC6715} \citep[143.06 km/s,][]{baumgardt19} by roughly the amount of the cluster velocity dispersion \citep[10.5 km/s,][2010 revision]{harris96}, and its metallicity is quite close to the one of the cluster, its distance is vastly larger than the estimated NGC6715 tidal radius \citep[7.5',][]{trager95}, so it is unlikely that the star originated within NGC6715.

This work is mainly based on the analysis of the UVES-slit spectra of J1907, observed in two 3005s exposures starting on 2009-04-25, 07:28:27 and 08:25:22 UT. The spectra were obtained in Dichroic 1 mode, with central wavelengths of 390nm and 580nm in UVES blue and red arm respectively, at a low airmass (1.13 - 1.01). The slit was set to 1.4'' on both arms, but seeing conditions were excellent ($\sim$ 0.45"), thus increasing the resolution to about R=60000. The present analysis is based on science-ready reduced spectra released by ESO\footnote{\url{http://archive.eso.org/wdb/wdb/adp/phase3_spectral/form}}. Coadding the two spectra a S/N$\sim$80 per sample (1x1 binning) around 630nm is reached. 

   \begin{figure*}
            \includegraphics[width=\hsize]{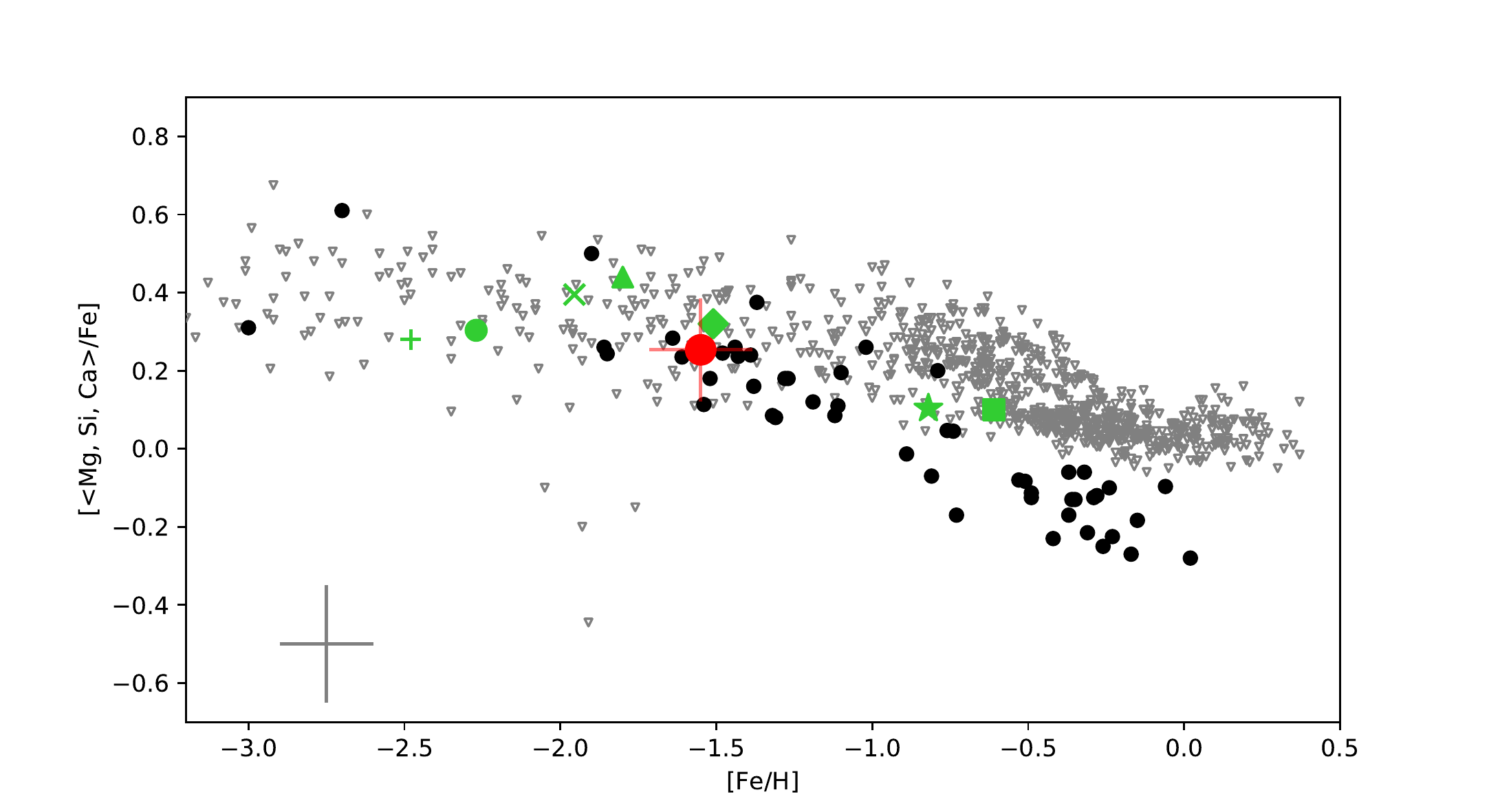}
      \caption{\alphafe~ vs. \feh~ for J1907 (large red circle with error bars), compared to relevant samples. Small open grey triangles, MW stars \citep{venn04,reddy06}. Black filled circles, \sgr~ main body stars \citep[][Sbordone et al., 2020 in prep.]{monaco05,sbordone07,hansen18}. Green large symbols, average values for globular clusters associated with \sgr: square, \object{Ter 7}; star, \object{Pal 12}; diamond, \object{NGC 6715}; triangle, \object{Arp 2}; X, \object{NGC 5634}; circle, \object{Ter 8}; cross, \object{NGC 5053} \citep{sbordone05b,cohen04,carretta10,mottini08,carretta14,sbordone15}. \alphafe~ is here defined as the simple average of [Mg/Fe], [Si/Fe], and [Ca/Fe], or any subsample of them available for each specific star. A reference error bar of $\pm 0.15$ dex on both axis is also plotted.}
         \label{alphafe}
   \end{figure*}

\begin{table}
\caption{Coordinates and parameters for J1907. Internal parameters errors are quoted (see Appendix \ref{paramerrs}). Magnitudes and coordinates come from \citet{giuffrida10}, proper motions have been derived from Gaia DR2}
\label{tableparams}   
\centering                      
\begin{tabular}{c c}  
\hline
GIU J190734.24-315102.1 &  \\    
\hline                        
RA (J2000)  & 19h 07m 34.24s       \\
DEC (J2000) & -31deg 51' 02.05''   \\
Mag (V, I)  & 17.742, 15.466       \\
V$_{\mathrm{rad}}$ & 155.04 $\pm$ 0.35 km/s \\
pm (RA$*$, DEC) & -2.65$\pm$ 0.13, -1.32$\pm$ 0.12 mas/yr \\
\Teff & 4753$^{+180} _{-160}$ K  \\
\logg & 1.75 $\pm 0.1$ cm s$^{-2}$ \\
\Vturb & 1.51 $\pm 0.2$  km/s\\
\feh & -1.55 $\pm 0.17$\\
\hline                                
\end{tabular}
\end{table}

An additional spectrum was obtained on August 21, 2015 (23:22:55 UT) with the MIKE spectrograph at Magellan Clay telescope in Las Campanas observatory, under Chilean time. The single 1200s spectrum covers the 483-916nm range and was taken with the 0.7''x5'' slit in mediocre transparency conditions, for the purpose of looking for radial velocity variations, and has S/N$\sim$15 per sample (2x2 binning) around 650nm. 

\begin{figure}
   \centering
   \includegraphics[width=\hsize]{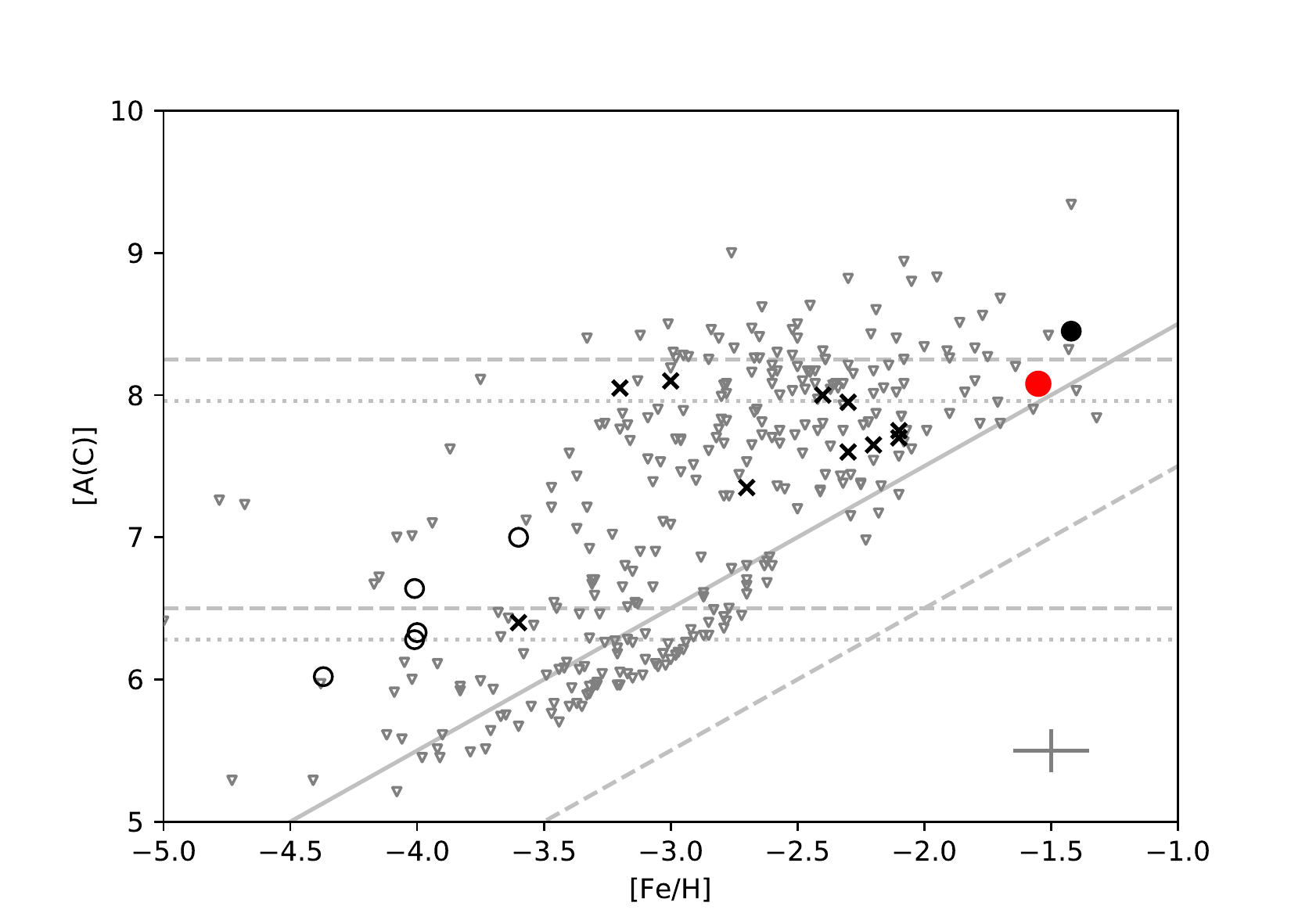}
      \caption{A(C) vs. [Fe/H] for J1907 (red filled circle) vs. different literature samples from \citet[][grey triangles]{yoon16} , \citet[][black  X]{hansen19}, \citet[][black open circles]{bonifacio18}, star \object{SDSS J100714+160154} from \citet[][black filled circle]{frebel14}. Horizontal lines describe the high- and low-carbon band according to \citet[][dashed]{spite12}, or \citet[][dotted]{yoon16}. Diagonal lines are [C/Fe]=0 (dashed) and +1.0 (continuous). All comparison samples, except SDSS J100714+160154, belong to the MW.
              }
         \label{theCplot}
\end{figure}

Radial velocities were determined by cross-correlating the spectra against a synthetic template of similar atmospheric parameters, employing the red arm spectra in the UVES case. The two UVES spectra gave 155.0 and 154.6 $\pm 0.7$ km/s, respectively. The MIKE spectra yield  155.3 $\pm 0.5$ km/s. As a consequence, there is no evidence that the star shows significant radial velocity variations. In Table\,\ref{tableparams} the weighted mean of the three measurements is given as reference.
The radial velocity is compatible with a membership with Sgr. \citet{bellazzini08} derives a radial velocity for the {\em core} of \sgr~ (up to 9' from the center) of 141 km/s, with a dispersion of roughly 10 km/s. \citet{majewski13}, covering larger distances from the center (up to about 2.5 degrees) detect a significant increase in velocity dispersion with increasing distance from the galaxy core, more pronounced for the more metal poor population. J1907 is even farther away from the \sgr~ center, at the extreme of the covered range along the galaxy major axis, and its velocity is well compatible with the other \sgr~ members observed in the same field (Sbordone et al., 2020, in prep.).
Proper motions, as derived from Gaia DR2, also confirm a membership in \sgr: J1907 shows $\mu$=-2.654,-1.323 (RA, DEC, mas/year), versus a \sgr~ average of -2.692,-1.359 \citep{gaia18c}. Since \sgr~ proper motions are very tightly clustered, J1907 can be considered a {\it bona fide} member of \sgr.

\section{Parameters determination and abundance analysis}
\label{params_and_abus}

Stellar parameters (listed in Table\,\ref{tableparams}) were determined spectroscopically by means of the \mygi~ automated abundance analysis code \citep{sbordone14}. 
The employed grid of synthetic spectra was the same used in \citet{sbordone15}, and was based on ATLAS12 1D-LTE atmospheric models and SYNTHE synthetic spectra \citep{kurucz05,sbordone04,sbordone05,castelli05}. The input list of features passed to \mygi~ was also used in \citet{sbordone15}, but the code may have kept a different subset of them. Due to the high C abundance, and because \mygi~ is not designed to derive precise abundances for elements strongly deviating from a solar-scaled composition \citep[see][for details]{sbordone14}, a number of abundances (C, N, O, plus n-capture elements) were derived fitting relevant features manually. 

\begin{table}
\caption{Atomic data and [X/Fe] for lines measured manually.}             
\label{manual_abus}    

\centering          
\begin{tabular}{l c c c }     
\hline\hline       
Wavelength& lower e.  & \loggf &  [X/Fe] \\
(nm)      & (eV)      &        &        \\
\hline
\ion{Sr}{ii} & & & \\
\hline
4077.714  & 0.000   &    0.15  &  0.47 \\    
\hline
\ion{Y}{ii} & & & \\
\hline
485.4867  &  0.992  &  -0.380  &  0.55   \\  
488.3684  &  1.083  &   0.070  &  0.45   \\  
520.5731  &  1.032  &  -0.340  &  0.40   \\  
\hline
\ion{Zr}{i} & & & \\
\hline
535.0851  &   2.322 &  -0.590  &  0.95   \\  
\hline
\ion{Zr}{ii} & & & \\
\hline
496.2310  &   0.971 &  -2.000  &  0.90   \\  
\hline
\ion{Ba}{ii} & & & \\
\hline
585.3686  &   0.604 &  -2.066  &  1.40   \\  
\hline
\ion{La}{ii} & & & \\
\hline
4970.383  &  0.321  &  -1.683  &  1.30   \\  
4986.765  &  0.173  &  -2.300  &  1.25   \\  
5259.380  &  0.173  &  -1.950  &  1.30   \\  
\hline
\ion{Ce}{i} & & & \\
\hline
524.9605  &  0.410  &  -0.630  &  1.40   \\   
\hline
\ion{Ce}{ii} & & & \\
\hline
488.2463  &  1.527  &   0.190  &  1.30   \\   
489.3952  &  1.326  &  -0.538  &  1.30   \\   
491.4924  &  0.879  &  -0.810  &  1.15   \\   
518.7458  &  1.211  &   0.170  &  1.30   \\   
523.7067  &  1.319  &  -0.620  &  1.30   \\   
\hline
\ion{Pr}{ii} & & & \\
\hline
525.9614  &  0.633  &  -3.727  &  1.30   \\   
532.2710  &  0.482  &  -1.878  &  1.20   \\   
538.1260  &  0.508  &  -0.461  &  1.20   \\   
\hline
\ion{Nd}{ii} & & & \\
\hline
406.1080  &  0.471  &   0.550  &   1.30  \\   
485.9030  &  0.320  &  -0.440  &   1.50  \\   
487.6110  &  0.559  &  -1.230  &   1.30  \\   
488.2880  &  0.742  &  -1.410  &   1.20  \\   
494.2960  &  0.742  &  -1.130  &   1.30  \\   
516.7920  &  0.559  &  -1.180  &   1.30  \\   
517.6780  &  1.120  &  -0.840  &   1.30  \\   
518.1169  &  0.859  &  -0.600  &   1.17  \\   
525.5510  &  0.204  &  -0.670  &   1.50  \\   
\hline
\ion{Sm}{ii} &&& \\
\hline
4913.260  &   0.659   &  -0.93 &  1.28 \\ 
4948.630  &   0.543   &  -0.95 &  1.30 \\ 
4952.370  &   0.333   &  -1.25 &  1.06 \\ 
\hline
\ion{Gd}{ii} & & & \\
\hline
517.6290  &  1.059  &  -0.710  &   1.15  \\   
\hline
\ion{Dy}{ii} & & & \\
\hline
516.9690  &  0.100  &  -1.950  &   1.20  \\  
\hline
\end{tabular}

\end{table}

Abundances for \ion{Na}{i}, \ion{Mg}{i}, \ion{Si}{i}, \ion{Ca}{i}, \ion{Sc}{ii}, \ion{Ti}{i}, \ion{Ti}{ii}, \ion{Cr}{i}, \ion{Mn}{i}, \ion{Fe}{i}, \ion{Fe}{ii}, \ion{Co}{i}, \ion{Ni}{i}, and \ion{Zn}{i} were obtained by \mygi. Features for \ion{C}{i}, \ion{N}{i}, \ion{O}{i}, and \ion{Eu}{ii} were fitted by means of FitProfile \citep{thygesen16}. All the remaining abundances were derived using MOOG \citep[][version 2014]{MOOG}.

Table \ref{manual_abus} lists the  features used for the abundances measured with MOOG. Following Appendix A of \citet{sbordone15}, details of the synthetic fits for all the lines measured by \mygi~ are made available online. Given the close affinity between \mygi~ and FitProfile, the lines fitted with the latter have been included as well.

\begin{itemize}

\item {\bf Carbon, Nitrogen and Oxygen:} The carbon abundance was derived by fitting the CH G-band. From the range
between 422.9\,nm and 423.2\,nm, we also derived $^{12}$C/$^{13}$C. The nitrogen abundance was derived by fitting four CN features around 648nm, accounting for the C and O abundances in the synthesis. The oxygen abundance was derived from the 630nm [\ion{O}{i}] line. The blue wing of the line is marginally affected by a telluric line that does not compromise the fit, as it has been verified by means of Molecfit \citep{smette15,kausch15}.

  \begin{figure}
  \centering
  \includegraphics[width=\hsize]{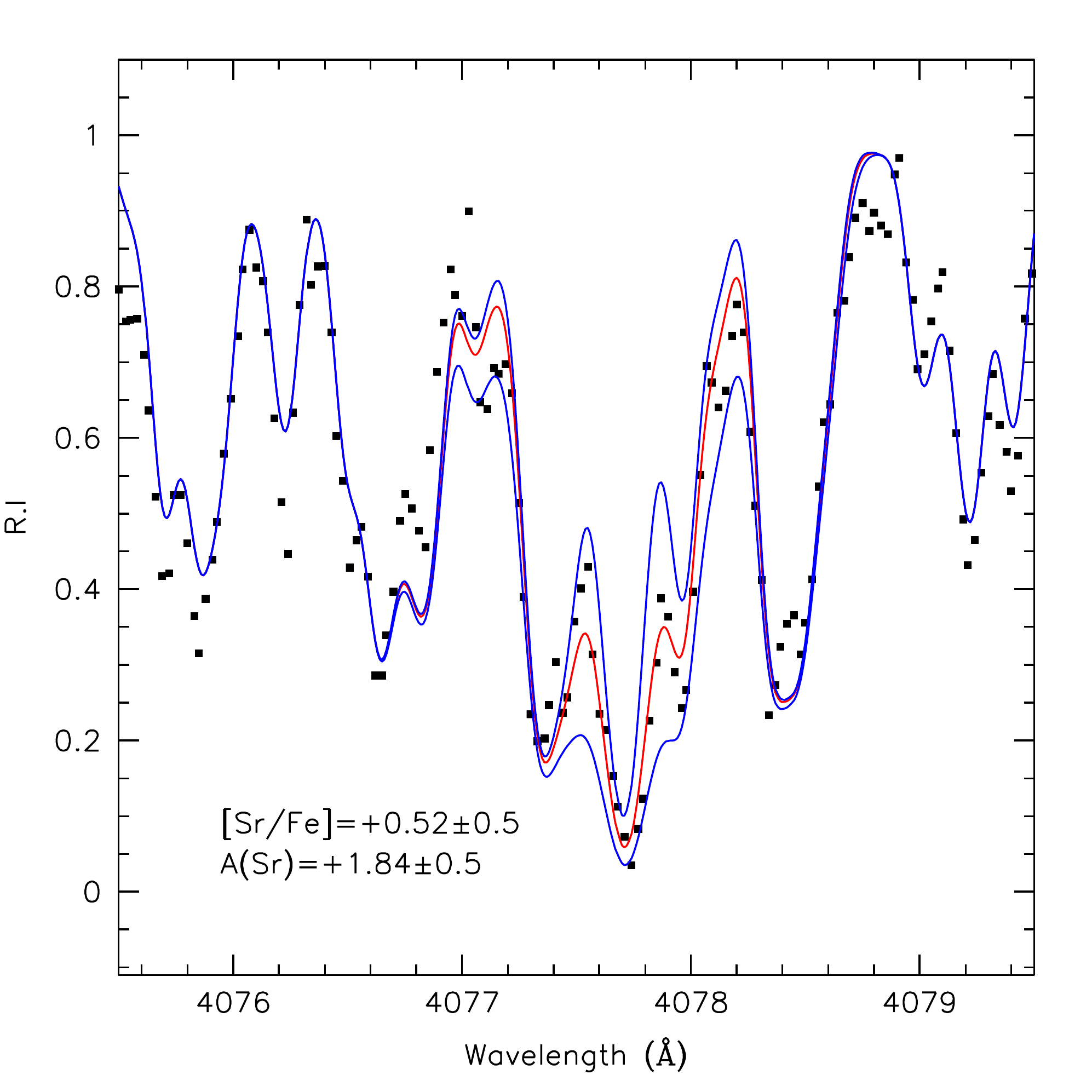}
      \caption{Fit of the 407.77 nm \ion{Sr}{ii} feature: The best fit, A(Sr)=1.84$\pm$0.5 is represented by the red curve, while the two blue lines represent models corresponding to $\pm$0.5 dex.
              }
         \label{strontium}
  \end{figure}

\item {\bf Strontium:} the only viable \ion{Sr}{ii} line in the spectrum is at 407.77 nm, and is extremely strong, to the point that it exceeds saturation and enters the damping part of the curve of growth. 
However, the wings are strongly blended. A manual fit, shown in Fig. \ref{strontium}, allows to derive the abundance quoted here, but we do not trust  the estimate to better than $\pm$0.5 dex (the abundance difference of the two fits plotted around the best one in Fig. \ref{strontium}). 

\item {\bf Europium:} three \ion{Eu}{ii} lines were analyzed in J1907, for all we adopted the \citet{lawler01} atomic data, hyperfine and isotopic splitting. The bluest, at 412.960 nm, is heavily saturated and strongly blended. It provides a tentative fit at [Eu/Fe]=+0.7. The 643.764 nm line is much weaker but provides an excellent fit at [Eu/Fe]=+0.97. This line is blended with a potentially problematic \ion{Si}{i} line at 643.770nm. \citet{jacobson13} derived an astrophysical \loggf~ of -2.3 for this line. In our linelist the transition is included with a higher \loggf~ (-2.053), but even so, due to the low metallicity of the star, the line is extremely weak and does not affect the Eu line. The 664.506 nm line, finally is one of the most frequently used to determine Eu abundances, and appear isolated and well fitted, giving [Eu/Fe]=+1.08. Due to the high uncertainty on \ion{Eu}{ii} 412.960nm, we rejected that line and used the transitions at 643 and 664 nm only.

\end{itemize}

\begin{table}
\caption{Abundances for J1907. See Sect. \ref{params_and_abus} for details.  Error for species with a single measured line are estimated according to Appendix \ref{paramerrs}.}            
\label{tableabunds}      

\centering          
\begin{tabular}{l r c r r c r c }     
\hline\hline       
Ion          & N$_{\mathrm{lin}}$ & A(X)$_{\odot}$ & A(X) & [X/H] & $\pm$ & [X/Fe] & $\pm$ \\
\hline                    
\ion{C}{i}   &  1  &   8.50  &  8.08 & -0.42  & 0.17&  1.13 &0.24\\
\ion{N}{i}   &  4  &   7.86  &  6.82 & -1.04  & 0.21  &  0.51 & 0.26 \\
\ion{O}{i}   &  1  &   8.76  &  7.61 & -1.15  & 0.17&  0.40 &0.24\\
\ion{Na}{i}  &  1  &   6.30  &  4.37 & -1.93  & 0.17& -0.38 &0.24\\
\ion{Mg}{i}  &  1  &   7.54  &  6.28 & -1.26  & 0.17&  0.29 &0.24\\
\ion{Si}{i}  &  2  &   7.52  &  6.12 & -1.40  & 0.08  &  0.15 & 0.18 \\
\ion{Ca}{i}  &  9  &   6.33  &  5.09 & -1.24  & 0.16  &  0.32 & 0.23 \\
\ion{Sc}{ii} &  7  &   3.10  &  1.70 & -1.40  & 0.16  &  0.15 & 0.23 \\
\ion{Ti}{i}  &  6  &   4.90  &  3.60 & -1.30  & 0.18  &  0.25 & 0.24 \\
\ion{Ti}{ii} &  3  &   4.90  &  3.44 & -1.46  & 0.48  &  0.09 & 0.51 \\
\ion{Cr}{i}  &  4  &   5.64  &  4.04 & -1.60  & 0.23  & -0.05 & 0.28 \\
\ion{Mn}{i}  &  6  &   5.37  &  3.62 & -1.75  & 0.50  & -0.19 & 0.52 \\
\ion{Fe}{i}  & 51  &   7.52  &  5.97 & -1.55  & 0.17  &    -- &   -- \\
\ion{Fe}{ii} & 10  &   7.52  &  5.97 & -1.55  & 0.16  &    -- &   -- \\
\ion{Co}{i}  &  1  &   4.92  &  3.19 & -1.73  & 0.17& -0.18 &0.24\\
\ion{Ni}{i}  &  2  &   6.23  &  4.67 & -1.56  & 0.22  & -0.01 & 0.28 \\
\ion{Zn}{i}  &  1  &   4.62  &  3.87 & -0.75  & 0.17&  0.80 &0.24\\
\ion{Sr}{ii} &  1  &   2.92  &  1.84 & -1.08  & 0.50  &  0.47 & 0.52 \\
\ion{Y}{ii}  &  3  &   2.21  &  1.13 & -1.08  &   --  &  0.47 & 0.08 \\
\ion{Zr}{i}  &  1  &   2.58  &  1.95 & -0.63  & 0.17&  0.95 &0.24\\
\ion{Zr}{ii} &  1  &   2.58  &  1.95 & -0.63  & 0.17&  0.90 &0.23\\
\ion{Ba}{ii} &  1  &   2.18  &  2.03 & -0.15  & 0.17&  1.40 &0.23\\
\ion{La}{ii} &  3  &   1.10  &  0.83 & -0.27  & 0.17&  1.28 &0.23\\
\ion{Ce}{i}  &  1  &   1.58  &  1.32 & -0.26  & 0.17&  1.40 &0.24\\
\ion{Ce}{ii} &  5  &   1.58  &  1.32 & -0.26  &   --  &  1.27 & 0.06 \\
\ion{Pr}{ii} &  3  &   0.72  &  0.40 & -0.32  &   --  &  1.23 & 0.06 \\
\ion{Nd}{ii} &  9  &   1.42  &  1.19 & -0.23  &   --  &  1.32 & 0.11 \\
\ion{Sm}{ii}&  3  &   1.00  &  0.66 & -0.34  & 0.13  &  1.21 & 0.20 \\
\ion{Eu}{ii} &  2  &   0.52  &  0.00 & -0.52  & 0.08  &  1.03 & 0.17 \\
\ion{Gd}{ii} &  1  &   1.07  &  0.67 & -0.40  & 0.17&  1.15 &0.23\\
\ion{Dy}{ii} &  1  &   1.10  &  0.75 & -0.35  & 0.17&  1.20 &0.23\\
\hline                  
\end{tabular}

\end{table}

   \begin{figure}
   \centering
   \includegraphics[width=\hsize]{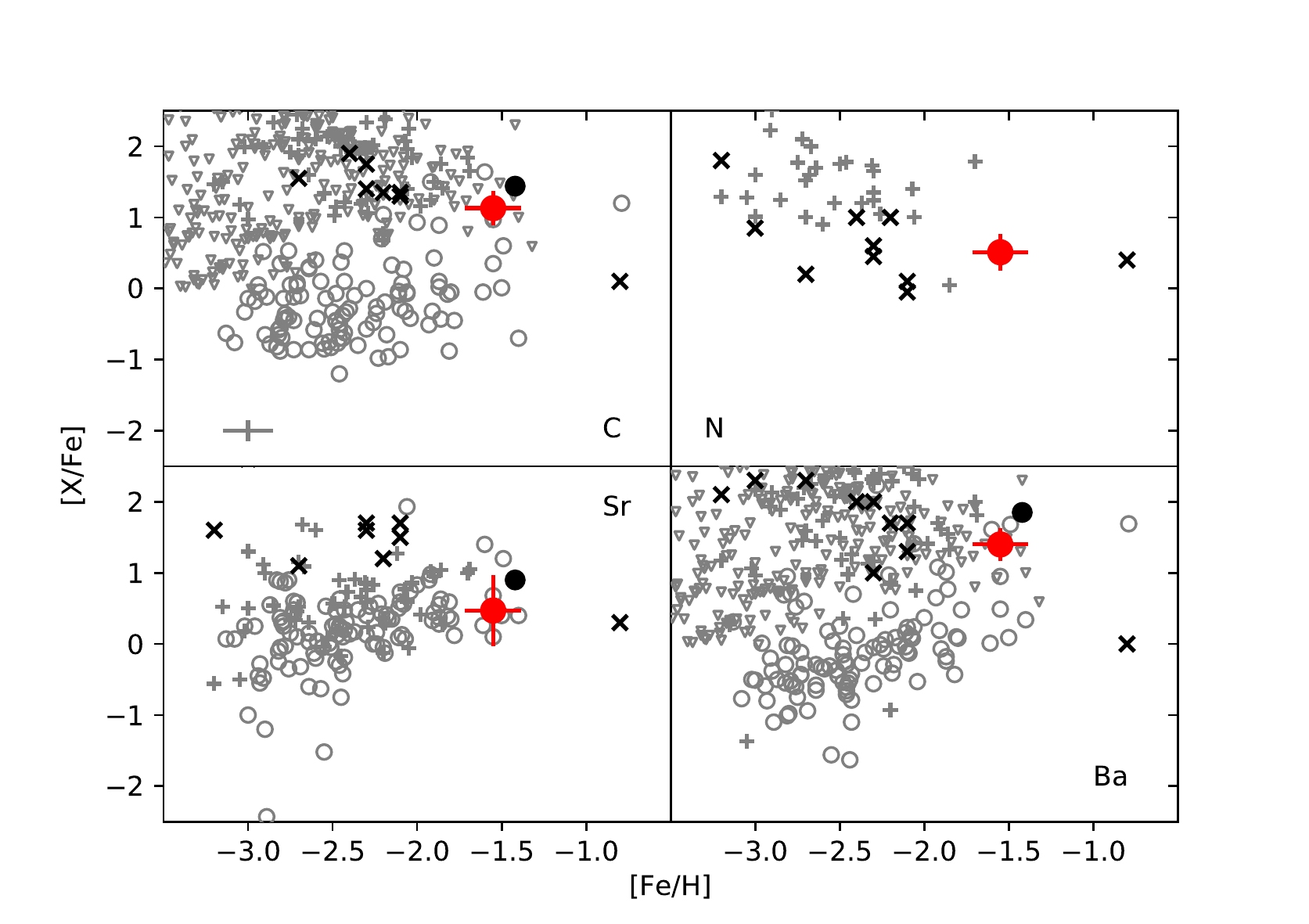}
      \caption{[X/Fe] vs. [Fe/H] for C, N, Sr, and Ba.  J1907 (red filled circle), \citet[][grey triangles]{yoon16}, \citet[][black X]{hansen19}, \citet[][grey open circles]{hansenT18}, \citet[][CEMP-s, -r, -s/r stars only, grey plus signs]{allen12}, star \object{SDSS J100714+160154} from \citet[][black filled circle]{frebel14}. A 0.15 dex error cross is added for reference. All comparison samples, except SDSS J100714+160154, belong to the MW.
              }
         \label{4plots}
   \end{figure}

In Table\,\ref{tableabunds} the final abundances are listed, together with the assumed solar abundances and [X/H] and [X/Fe] ratios. This set of solar abundances is that of \citet{caffau11} or \citet{lodders} for all elements
not present in \citet{caffau11}. 
Where multiple lines were used, the ``$\pm$'' columns list the $1\sigma$ dispersion around the average for [X/H], and the propagation accounting for the [\ion{Fe}{i}/H] or [\ion{Fe}{ii}/H] (for the [\ion{O}{i}] 630.03nm line, and ionized species) dispersion for the [X/Fe] values.
Together with the abundances listed in Table\,\ref{tableabunds}, we also derived a carbon isotopic ratio of $^{12}$C/$^{13}$C=12.

The $^{12}$C/$^{13}$C ratio is generally of difficult interpretation in CEMP-s and -r/s, where carbon enhancement is believed to be due to mass accretion of heavily C-enhanced ejecta from an AGB primary. AGB ejecta are believed to be very rich in $^{12}$C (the models used in this work have $^{12}$C/$^{13}$C comprised between 800 and 7600), while observed ratios in CEMP-s and -r/s are much lower (see \citealt{bisterzo11}, where values compiled from $\sim$ 50 stars span values of 4$\leq$$^{12}$C/$^{13}$C$\leq$90). This is due to the dilution of the AGB ejecta within the secundary convective envelope, which is less rich in $^{12}$C. However, a quantitative prediction is complicated by the unknown degree of dilution, the variations in $^{13}$C abundance depending on the secundary evolutionary stage, and the generally uncertain theoretical preditions on said $^{13}$C abundances, where extra-mixing seems to be required to fit observations \citep{busso10}.

   \begin{figure}
   \centering
   \includegraphics[width=\hsize]{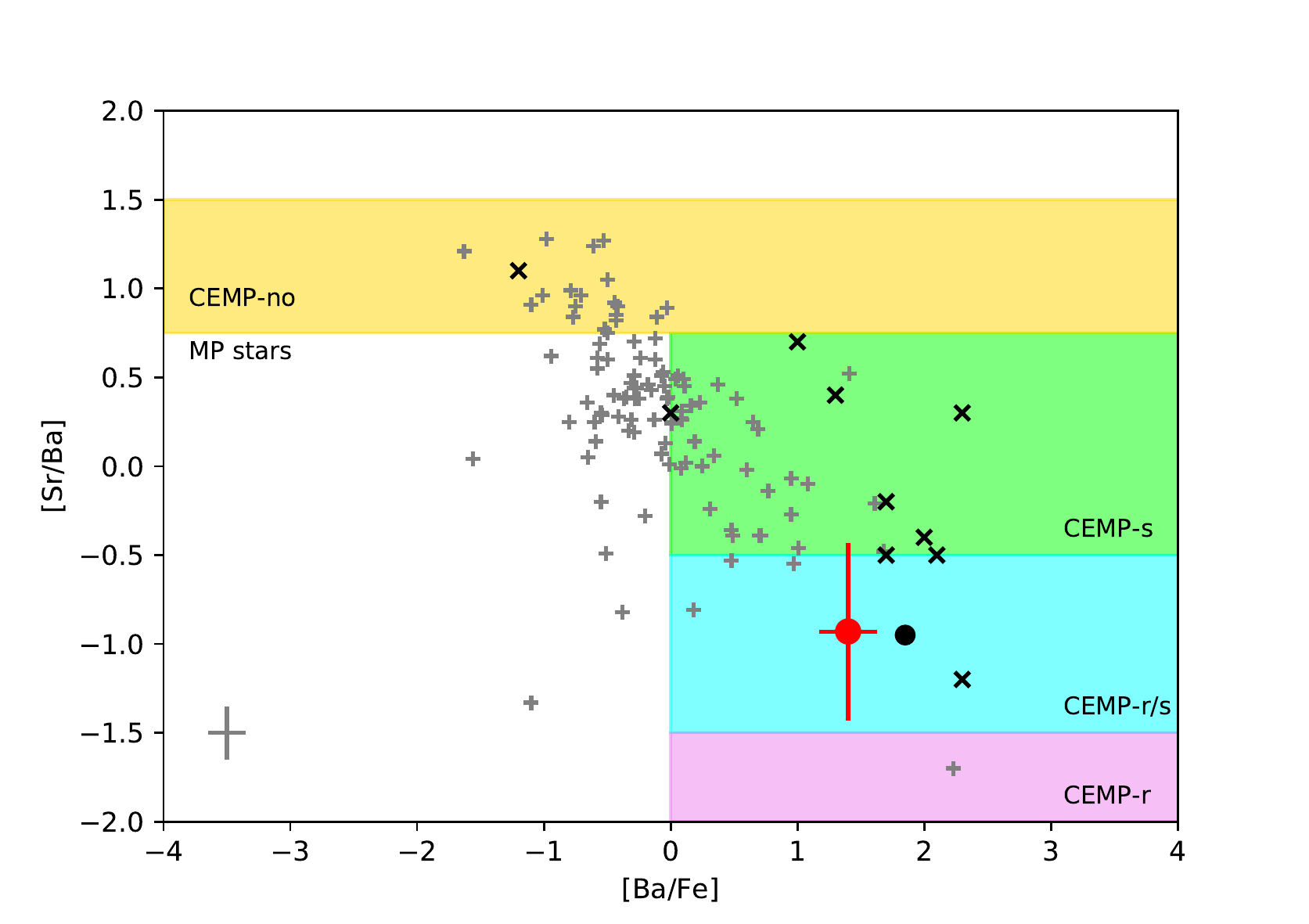}
      \caption{[Sr/Ba] vs. [Ba/Fe] for J1907 (red filled circle), \citet[][black X]{hansen19}, \citet[][grey crosses]{hansenT18}, star \object{SDSS J100714+160154} from \citet[][black filled circle]{frebel14}. The areas corresponding to the CEMP classification proposed in \citet{hansen19} are shaded and labeled. A 0.15 dex error cross is added for reference.
              }
         \label{SrBa}
   \end{figure}
  
\section{Discussion}

In Fig. \ref{alphafe} we plot \alphafe~ vs. \feh~ for J1907 , together with literature MW samples, other \sgr~ stars, and mean values for globular clusters associated with \sgr. In this respect, J1907 behaves as a typical member of the metal poor population that dominates the outskirts of \sgr: the very low \alphafe, and high \feh, observed in the central part of the galaxy  \citep{sbordone07,tolstoy09,mcwilliam13,hasselquist17} give way to higher levels of $\alpha$-enhancement, with \sgr~ being indistinguishable from the MW halo for metallicity below [Fe/H]$\sim$-2 \citep{sbordone15}. 

   \begin{figure}
   \centering
   \includegraphics[width=\hsize]{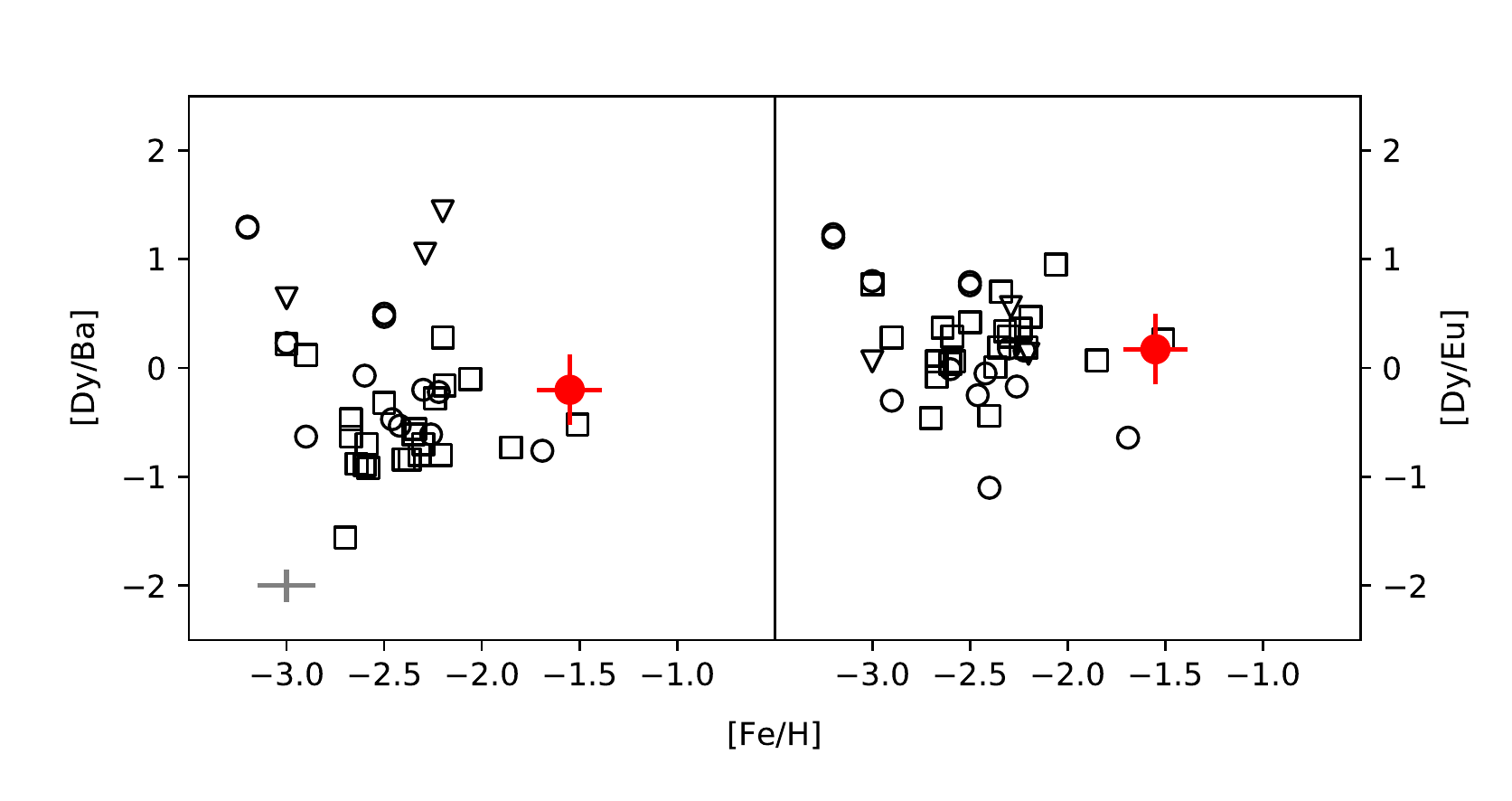}
      \caption{[Dy/Ba] and [Dy/Eu] plotted against [Fe/H] for J1907 (red filled circle) and  literature CEMP-r (triangles), CEMP-s (squares), and CEMP-rs (circles). A 0.15 dex error cross is added for reference.
              }
         \label{dysprosium}
   \end{figure}

However, the abundances of J1907 allow to classify it as a typical CEMP-r/s (or CEMP-i) star, and to the best of our knowledge it is the first star of this kind identified in \sgr.

      \begin{figure*}
            \includegraphics[width=\hsize]{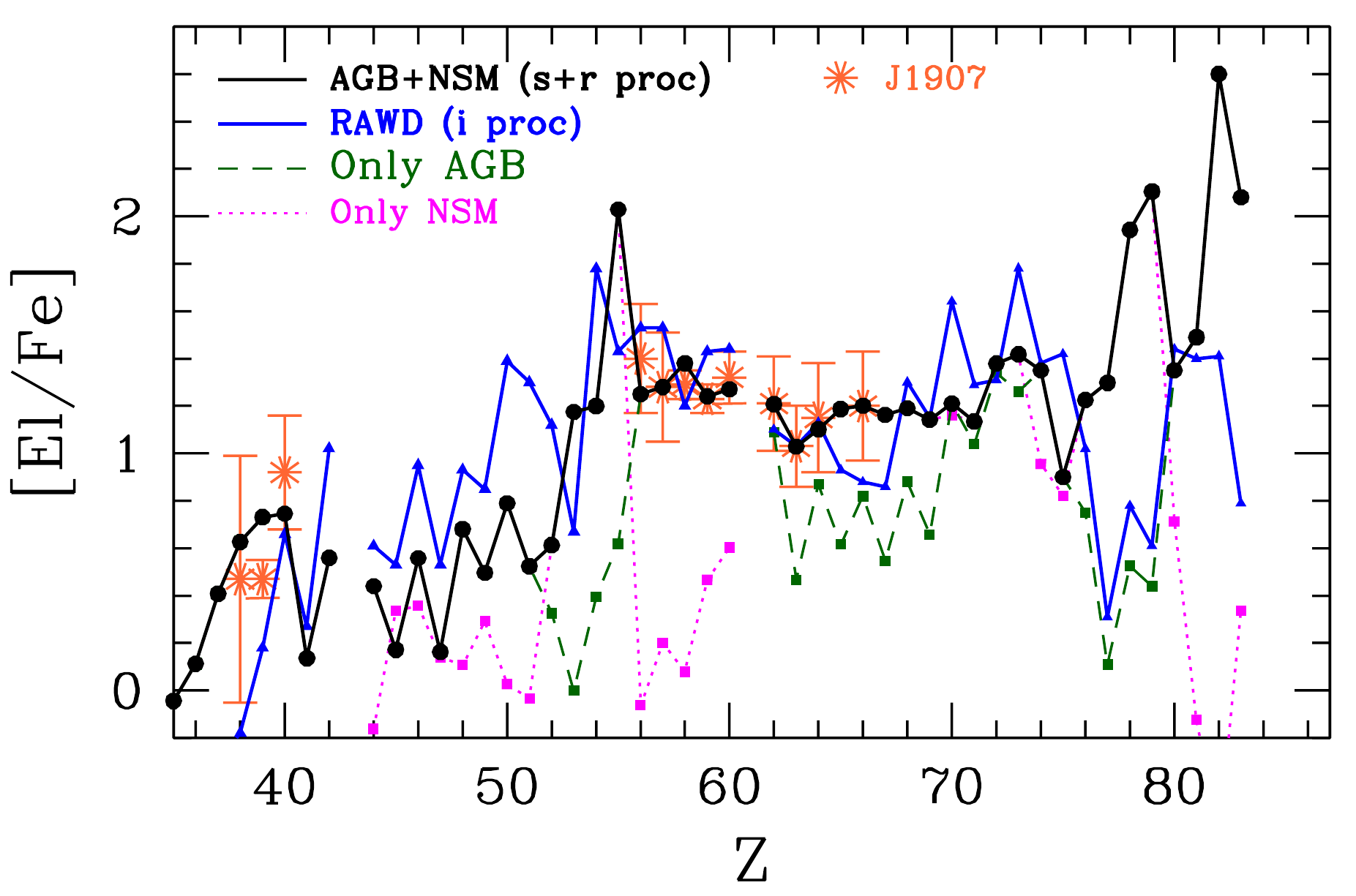}
      \caption{The abundance pattern of J1907 compared with yields from a NMS (magenta dotted curve), an AGB star (green dashed curve), a combination of them (black solid curve) and RAWDs (blue solid curve). See text for details.}
         \label{pattern_figure}
   \end{figure*}

The carbon abundance of J1907, plotted in Fig. \ref{theCplot} and \ref{4plots}, places it firmly in the ``high-carbon band'' defined by \citet{spite12} and \citet{bonifacio18} or the ``Group I'' defined by \citet{yoon16}.

Its [Sr/Ba] ratio also supports a CEMP-r/s classification according to \citet[][]{hansen19} (see Fig. \ref{SrBa}), and the same is true according to the \citet{beers05} classification scheme, with an [Eu/Fe]=+1.03, and consequently [Ba/Eu]=+0.37. Every measured element with atomic number $\geq$39 (Y) shows enhancements close to, or exceeding 1 dex with respect to iron ([X/Fe]).  

In Fig. \ref{dysprosium} we plot [Dy/Ba] and [Dy/Eu] against metallicity. Dysprosium abundances are rare in the literature. Out of the stars with n-capture enhancement (CEMP-r, -s, -rs) analyzed or collected from literature in \citet{allen12} a total of 19 have Dy abundances. To these we added 21 stars extracted from the SAGA database \citep{suda11}, and originally published in \citet{jonsell06}, \citet{behara10}, \citet{cui13}, \citet{placco13}, \citet{roederer14}, \citet{hansenT15}, \citet{hollek15}, \citet{placco15}, \citet{jorissen16}, \citet{gull18}.
%reference for what is what, apart from Allen:
%\object{RAVE J094921.8-161722} from \citet{gull18}, \object{CS 22881-036}, \object{CS 22879-029}, \object{CS 22896-136}, \object{CS 22945-024}, \object{CS 22947-187}, \object{CS 22956-102}, \object{CS 29513-014} from \citet{roederer14}, \object{HE 0017+0055} from \citet{jorissen16}, \object{HD201626} from , \object{HE 0054-2542}, \object{HE 0243-3044}, \object{HE 0440-3426} from \citet{hansenT15}, \object{HE 0414-0343} from \citet{hollek15}, \object{HE 1405-0822} from \citet{cui13}, \object{SDSS J0912+0216}, \object{SDSS J1036+1212}, \object{SDSS J1349-0229} from \citet{behara10}, \object{HE 0338-3945} from \citet{jonsell06}, \object{HE 2258-6358} from \citet{placco13}
It is common in the literature to study the [La/Eu] ratios of those stars to determine their relative s-process and/or r-process enrichments (see, e.g., \citealt{roede}). In order to perform such classification, we can alternatively use dysprosium surface abundances. In the left panel of Fig. \ref{dysprosium} it clearly emerges that CEMP-s stars have the lowest [Dy/Ba] values (and CEMP-r stars the highest ones). Moreover, we can further refine such analysis by comparing dysprosium and europium, which are both typical r-process elements (85\% and 95\% of their solar abundances, respectively; see \citealt{nikos2020}). Thus, the expected [Dy/Eu] spread between different classes of CEMP stars should be lower, as demonstrated in the right panel of Fig. \ref{dysprosium}. Both Fig. \ref{4plots} and \ref{dysprosium} do not show signs that the n-capture nucleosynthesis that affected J1907 is remarkably different from the one in stars belonging to the Milky Way.

\begin{figure}
            \includegraphics[width=\hsize]{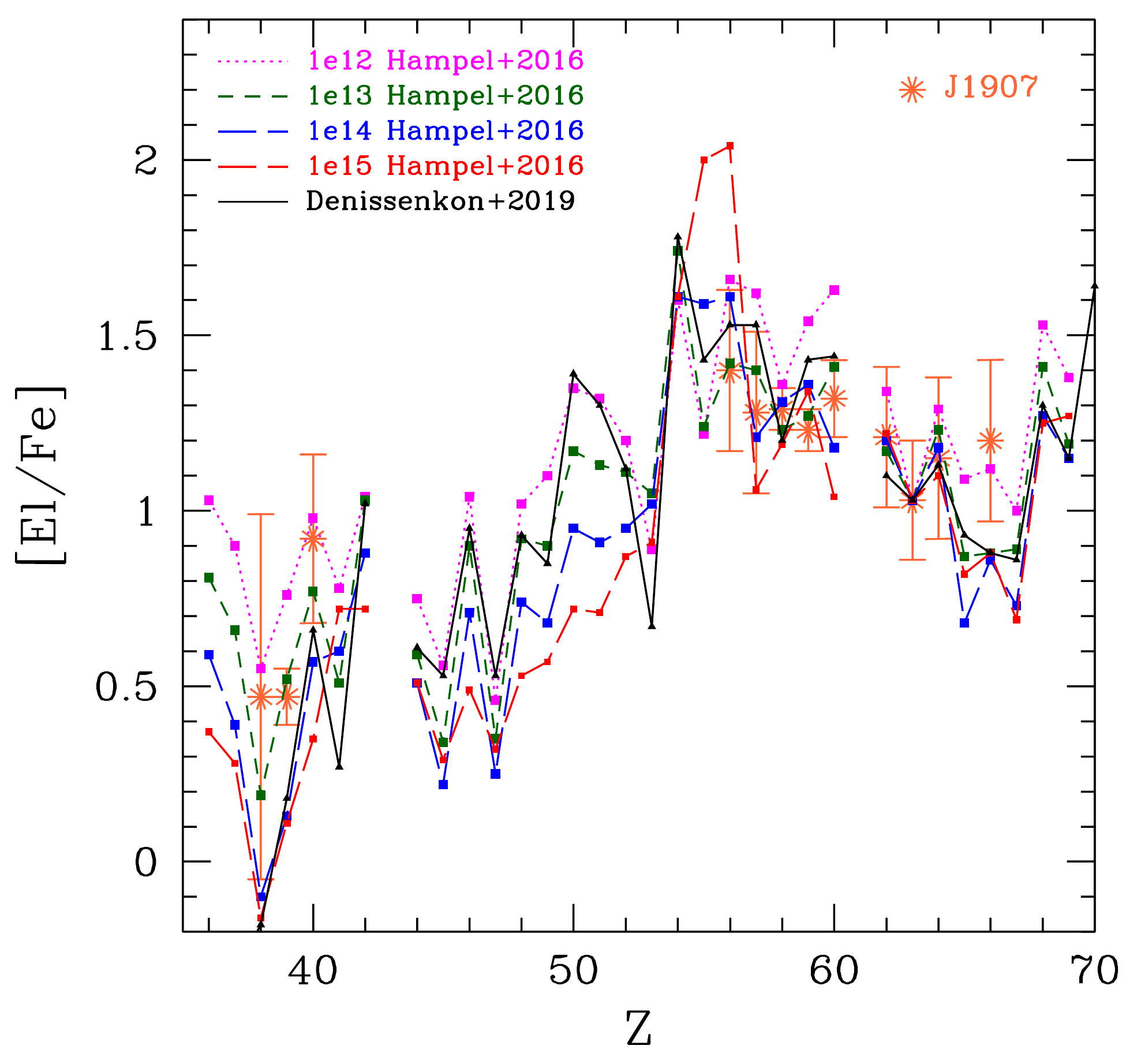}
      \caption{i-process nucleosynthesis calculation \citep{hampel16} for different neutron densities, compared to J1907. The distribution at [Fe/H]=-1.55 from \citet{Deni19} is also reported. See text for details.}
         \label{models_iproc}
   \end{figure}

In Fig. \ref{pattern_figure} we compare the abundance pattern in J1907 with a number of theoretical enrichment yields. 
In particular, we explore two completely different nucleosynthesis paths: a case with contributions from the slow and the rapid neutron capture processes (s-process and r-process; for reviews see \citealt{st06} and \citealt{Co19}) and a case characterized by the i-process only. In Fig. \ref{pattern_figure} we report different models:
\begin{enumerate}

    \item Magenta dotted curve: a Neutron Star Merger (NSM) final surface element distribution (derived from element yields obtained using the SKYNET code with $Y_e$=0.1;  \citealt{lip17}), normalized to the J1907 europium overabundance,  which is assumed as a representative element for the r-process.
    
    \item Dashed green curve: final surface element distribution of a low-mass low-metallicity AGB star (M=2 M$_\odot$, [Fe/H]=-1.67; \citealt{cri15}), normalized to the J1907 lanthanum overabundance, which is assumed as a representative element for the s-process.
    
    \item Dark solid curve: a combination of item 1 and 2, in which we assigned to each element the maximum abundance between the NSM and the AGB curves. Such a scenario sees in fact a pollution from two different sources: an AGB star providing the s-process enrichment and a NSM supplying elements typical of the r-process. Such an hypothesis has been largely explored in the past \citep{bisterzo10,bisterzo11,bisterzo12}. 
    
    \item Blue solid curve: i-process nucleosynthesis resulting from Rapidly Accreting White Dwarfs \citep{Deni19} with [Fe/H]=-1.55, normalized to the J1907 europium overabundance.
    
\end{enumerate}
Interpretation of the observed abundance pattern of J1907 is not straightforward, as it is often the case for CEMP-r/s stars.
Before proceeding with such analysis, however, some important caveats have to raised.\\
      \begin{figure}
            \includegraphics[width=\hsize]{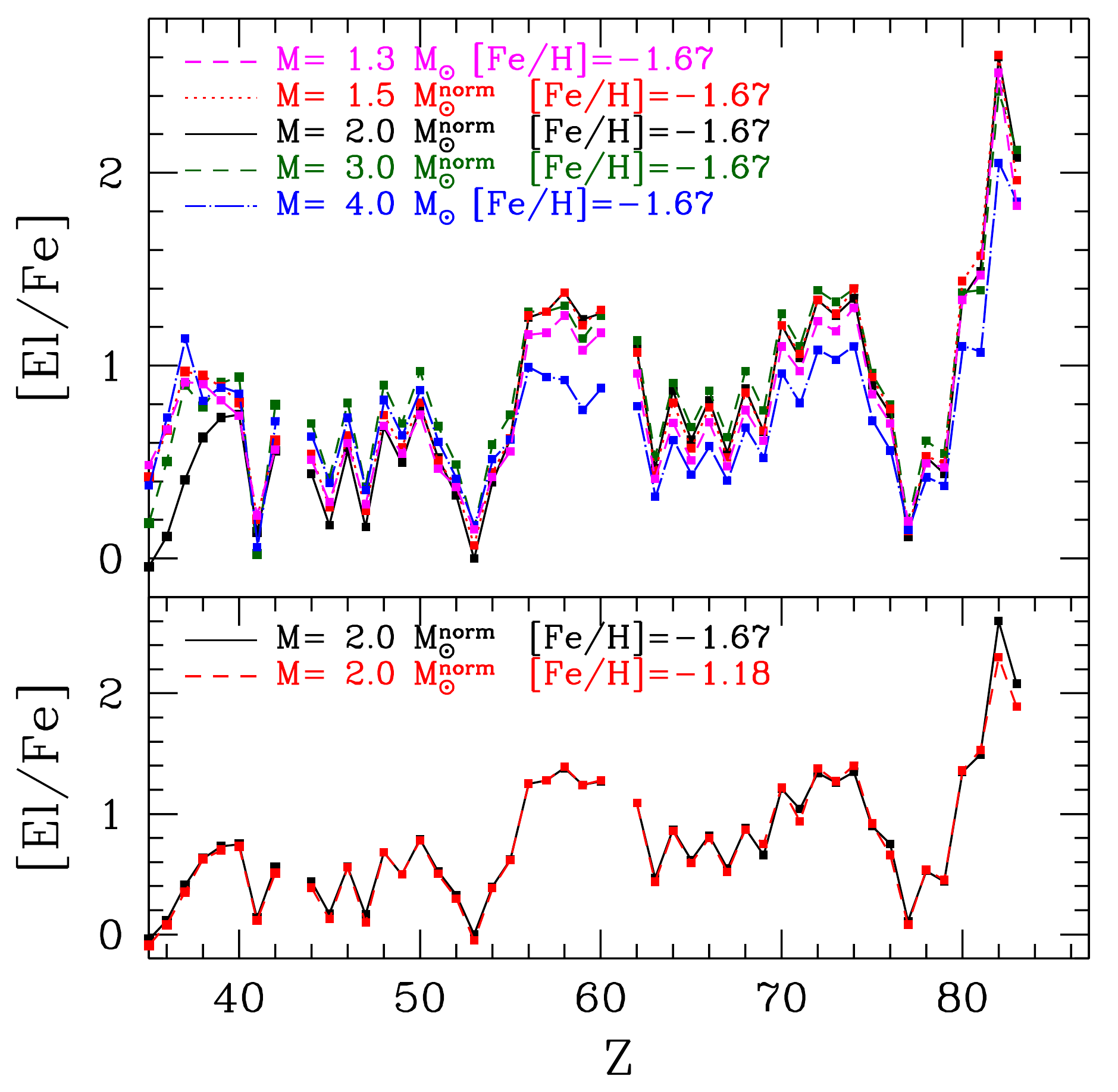}
      \caption{FRUITY AGB models with different initial masses (upper panel) and at different metallicities (lower panel). See text for details.}
         \label{models_figure}
   \end{figure}
Starting from AGB stars, it is worth to highlight that we find negligible variations by changing the metallicity (from [Fe/H]=-1.67 to [Fe/H]=-1.18, the nearest more metal-rich model available on the FRUITY database\footnote{http://fruity.oa-abruzzo.inaf.it/}; \citealt{cri15}) or the initial mass (1.5$\le $M/M$_\odot \le$3.0). This is proven by Figure \ref{models_figure}, where FRUITY models for different masses and metallicities are reported. Distributions with label "norm" have been normalized to the lanthanum abundance of J1907. Models with initial masses between 2 M$_\odot$ and 3 M$_\odot$ show almost the same distribution: this is due to the fact that, at these metallicities, both the light-s ({\it ls} elements: Sr-Y-Zr) and heavy-s ({\it hs} elements: Ba-La-Ce-Pr-Nd) elements are saturated. The only appreciable variations are found for lead, which has not been measured in our star (see later in the text). It has to be stressed that the 1.3 M$_\odot$ and the 4.0 M$_\odot$ models are unable to attain the enrichment level of heavy-s elements found in J1907. While for the 1.3 M$_\odot$ a slight increase in the mixing efficiency could  solve the problem, the situation for the 4.0 M$_\odot$ (and for more massive models) is different. In fact, a larger production of heavy elements would be compensated by a worsen fit to light-s elements. Thus, we can reasonably exclude massive AGBs from the polluters of J1907.
\\
    The high carbon content in a relatively metal rich star, the placement on the high-carbon-band, and the strong overabundance of s-process elements all point to J1907 having been affected by mass transfer from an AGB companion. The star does not show straightforward radial velocity variations, but the limited coverage and precision makes this merely an ``absence of evidence''. The high Eu, Gd, and Dy abundances, however,  are incompatible with an s-process-only source for neutron-capture elements.
    As for other CEMP-r/s stars, the abundances of Eu, Gd, and Dy can be explained with a pre-enrichment of the gas from which J1907 was formed. We hypothesize that such an enrichment comes from a NSM. The infrared re-brightening of the electromagnetic transient AT2017gfo \citep{pian2017}, following the 
    %LS:gravity%LS--
    gravitational waves event GW170817 \citep{ab17}, proved that NSM events are characterized by a rich r-process nucleosynthesis %LS:
    \citep{watson19}
    %LS--
    . A key parameter in those simulations is the electron-to-baryon ratio, the so-called Y$_e$. Depending on its initial value, completely different final distributions can be obtained (for a recent review see \citealt{Co19}). In our simulation we adopt a low Y$_e$, needed to develop a full nucleosynthesis up to the third r-process peak. For the sake of clarity, it has to be noted that different components have been identified in a single NSM event, each of them marked by a different Y$_e$. Moreover, additional r-process polluters have been suggested, as magneto-rotationally-driven SuperNovae \citep{nobuya} or magnetars \citep{siegel2019}: these stellar environments may in fact provide heavy element distributions similar to a NSM. A detailed comparison between those nucleosynthetic events is beyond the scope of our paper. We based our choice on the only (up-to-date) proven source for the r-process nucleosynthesis and thus we assume that a NSM event provided the r-process component of J1907.\\
    An alternative explanation to a combined pollution from the s-process and the r-process may come from the i-process. Such a process is thought to originate from the sudden mixing of hydrogen to very hot layers \citep{cr77}.
    One of the most uncertain ingredients characterizing i-process calculations is the stellar site hosting such a process. Up to date, three hypotheses have been explored with stellar evolutionary models:
    \begin{enumerate}
    \item {proton ingestions at the beginning of the thermally pulsing AGB phase of  very low-metallicity low-mass stars: this process appears remote for J1907, due to its relatively high metallicity, which prevents the convective protons to be mixed in He-rich layers (see \citealt{cristallo16} for details; see also \citealt{sta11} and \citealt{falk14} for very-low metallicity 3D hydrodynamical simulations). It has to be stressed that moderately metal-rich stars have been fitted with i-process calculations \citep{roe16,ko19}. However, these papers presented network calculations only, without clearly identifying the stellar site hosting such a process.}
    \item {proton ingestions at the end of the thermally pulsing AGB phase (the so-called Very Late Thermal Pulse): this process has been proposed to explain the peculiar nucleosynthesis of Sakurai's object \citep{falk11}. However, available calculations are for solar metallicity only.}
    \item {RAWDs, which are able to simultaneously produce both s-process elements (as lanthanum) and r-process elements (as europium): simulations for these events are available at different metallicities \citep{Deni19}.}
    \end{enumerate}
    In addition to these stellar models, network calculations have been published by \cite{hampel16}: even if performed at metallicities lower than J1907, they can be used to track its nucleosynthesis, due to the saturation effect of the ls and the hs component of the s-process (see above). In Figure \ref{models_iproc} we report network calculations for different neutron densities (normalized to europium) and we compare them to the J1907 distribution. From a  visual inspection, we can safely exclude the lowest ($10^{12}$ cm$^{\rm -3}$) and the highest ($10^{15}$ cm$^{\rm -3}$) neutron density cases. The first is not able to fit any of the hs elements, while the latter does not match ls elements and barium. Intermediate cases (i.e. between $10^{13}$ cm$^{\rm -3}$ and $10^{14}$ cm$^{\rm -3}$) provide a good enough fit, with the exclusion of gadolinium (this problem is shared with the calculation by \citealt{Deni19}, which also under-produces light-s elements). On the other hand, these network calculations provide a good fit to the large observed [Zr/Y], which is difficult to be obtained with s-process models.\\ 
    In the comparison of Figure \ref{pattern_figure}, we opt for discussing only distribution obtained with stellar evolutionary calculations, but in the future useful constraints can also be derived from the above-mentioned network calculations.

    For the sake of clarity, it has to be stressed that both scenarios (RAWD vs. AGB+NSM) have difficulties in explaining from a statistical point of view the CEMP populations observed in the halo of our Galaxy \citep[][even if this analysis was performed for lower metallicities]{abate16}.
    With the current element list it is objectively difficult to disentangle between the two described scenarios. The situation would be completely different if at least one element belonging to the couples Pt-Au and Pb-Bi would be detected. For those elements, in fact, the i-process scenario shows definitely lower enhancements with respect to the s+r scenario, where they result largely abundant (the Pt-Au couple due to a pre-existing r-process, while the Pb-Bi couple from AGB pollution). Further observations dedicated to this star would be tremendously useful. 

J1907 is fairly similar, from a chemical point of view, to SDSS J100714+160154, studied in the ultra-faint dwarf galaxy \object{Segue 1} by \citet{frebel14},
as can be seen in Fig. \ref{theCplot}, \ref{4plots}, \ref{SrBa}. However, SDSS J100714+160154 shows more pronounced s-process enrichment (e.g. [Sr/Fe]=0.9, [Zr/Fe]=1.4, [Ba/Fe]=1.85) and a measured Pb abundance, while Pb was not detectable in the available spectrum of J1906. The two stars share a high [hs/ls] ratio\footnote{This ratio defines the enrichment of the second peak of the s-process (hs elements) with respect to the first peak (ls elements) and it is defined as [hs/ls]=[hs/Fe]-[ls/Fe] in the usual spectroscopic notation.}. On the other hand, r-process elements are less enhanced, with [Eu/Fe]=0.7. This would make it a CEMP-s star as per the \citet{beers05} classification, while the \citet{hansen19} classification would indicate it as a CEMP-r/s (Fig. \ref{SrBa}).

   \section{Conclusions}
   
We present a chemical analysis for the star GIU J190734.24-315102.1 (J1907), a member of \sgr. The star is a moderately metal poor ([FeH]=-1.55) giant, displaying strong carbon enhancement ([C/Fe]=+1.13), and n-capture enhancement of both s- and r-process elements ([Ba/Fe]=+1.4, [Eu/Fe]=+1.0). 

To the best of our knowledge, J1907 is the first CEMP-r/s detected in the \sgr. The situation is less obvious in other dwarf galaxies of the Local Group.

\begin{itemize}
    \item \object{UMi K} \citep{shetrone01}, \object{Fornax 21} \citep{shetrone03}, and \object{Sculptor 982} \citep{geisler05} are the earliest reports we can find of Ba- and Eu- enhanced stars in Local Group dwarf Spheroidal galaxies. Their heavy element abundances would classify them as CEMP-r/s according to the \citet{beers05} scheme, but their carbon abundance was never determined. Sculptor 982 is reported \citep{hill19,skuladottir19} to show very strong CN bands, a further hint in this direction.
    \item The aforementioned SDSS J100714+160154 presented in \citet{frebel14} is borderline between CEMP-s and CEMP-r/s, to the extent a sharp threshold at [Ba/Eu]$=$0.5 makes physical sense.
    \item \citet{chiti18}, in a search for CEMP stars in the \object{Sculptor dSph}, identifies 11 candidate CEMP-s / r/s, based on their high C abundance, placing them in the ``high-carbon band'' (see also Fig. \ref{theCplot}). Only three of them have Ba abundances, and only one of these does indeed show a significant Ba overabundance. None have Sr or Eu abundances measured.
\end{itemize}

With the current available models, J1907 appears to be best explained as the product mass exchange from an AGB companion within a binary system pre-enriched at high concentration by the yields of a NS-NS merger. 

It seems unlikely that this r-enrichment event had only affected one star that thus also underwent AGB mass tranfer. Other r-enriched stars should be present among \sgr~ stars of similar metallicity, formed together with J1907. Given the star placement at the extreme of the \sgr~ main body major axis, a good number of such stars could have since been stripped by the interaction with the MW, and reside currently in the \sgr~ stream.

\begin{acknowledgements}
This paper is dedicated to the dear memory of the late Fiorella Castelli, 
whose work in the modeling and analysis of stellar atmospheres has been a fundamental guidance for us.
The authors thank Diego Vescovi for help in the use of the SKYNET code.
EC and PB gratefully acknowledge support from the French National Research
Agency (ANR) funded project ``Pristine'' (ANR-18-CE31-0017).
EC and PB are thankful to ESO - Santiago for hosting them during the
preparation of this manuscript.
SV gratefully acknowledges the support provided by Fondecyt reg. n. 1170518.
This work has made use of data from the European Space Agency (ESA) mission
{\it Gaia} (\url{https://www.cosmos.esa.int/gaia}), processed by the {\it Gaia}
Data Processing and Analysis Consortium (DPAC,
\url{https://www.cosmos.esa.int/web/gaia/dpac/consortium}). 

This work has made use of data from APOGEE, part of SDSS IV. 
 The SDSS web site is www.sdss.org.

\end{acknowledgements}

\bibliographystyle{aa} % style aa.bst
\bibliography{biblio} 

\begin{appendix}
\section{Estimation of errors on atmosphere parameters and abundances}
\label{paramerrs}
Table \ref{tableparams} quotes \mygi~ internal parameters errors. These are computed as follows. 
\begin{itemize}
    \item During \Teff~ calculation, the linear fit of the relationship between lower energy and and abundance given by individual \ion{Fe}{i} lines is computed, and \Teff~ iteratively adjusted until its slope is (very close to) zero \citep[see][]{sbordone14}. This populates a list of \Teff-slope couples, against which a second-order polynomial is fit. When the final \Teff~ is chosen, the error on the slope for that last fit is used together with this second-order polynomial to determine which temperatures would correspond to slopes equal to {\tt final\_slope$-$slope\_error} and {\tt final\_slope$+$slope\_error}.
    \item Similarly, \Vturb~ is determined by iteratively finding the value that zeroes the slope of the relationship between reduced equivalent width and abundance of individual \ion{Fe}{i} lines. As the final value is found, the error of the slope and the stored \Vturb / slope couples are used to determine the \Vturb~ uncertainty.
    \item Gravity is determined by enforcing equal average abundance to be given by \ion{Fe}{i} and \ion{Fe}{ii} lines. The internal error (line-to-line dispersion) of \ion{Fe}{i} and \ion{Fe}{ii} average abundances are summed in quadrature, and the uncertainty in gravity is determined by deriving the gravity that produces A(\ion{Fe}{i}) -- A(\ion{Fe}{ii}) offsets equal to {\tt final\_offset $-$ quadrature\_error} and {\tt final\_offset $+$ quadrature\_error}.
\end{itemize}
Given that the formal values of these uncertainties are sensitive to the way parameters changed during the actual parameter-search iteration, we conservatively round uncertainties to 10K, 0.1 km/s, and 0.1 cm s$^{-2}$ for \Teff, \Vturb, and \logg, respectively.

Quoted errors on abundances are estimated in different fashions.
\begin{itemize}
    \item For abundances of species measured with \mygi~ and based on at least two lines, uncertainties on [X/H] are line-to-line scatter, while the [X/Fe] values are summed in quadrature with either \ion{Fe}{i} or \ion{Fe}{ii} line-to-line scatter \citep[see][for details]{sbordone14}. \ion{N}{i} abundance, although determined with FitProfile, is based on multiple features and treated the same way.
    \item Species measured with MOOG and based on multiple features also quote line-to-line scatter, however, since in this case [X/Fe] was measured directly, only the scatter on it is reported.
    \item Species based on a single feature (with the exception of \ion{Sr}{ii}, see above) do not allow to directly probe the main sources of uncertainty. For isolated lines formal measurement uncertainty on the equivalent width, according to \citet{cayrel88} does not exceed 5-10\% for unsaturated lines at the typical S/N of this spectrum, with a negligible \citep[0.04 dex, see also][table 3 for a Montecarlo test on simulated, noise-injected spectra]{caffau13} impact on abundances. Line-to-line scatter in the best sampled species (\ion{Fe}{i} and \ion{Fe}{ii}) is around 0.16 dex, even on carefully selected lines, underscoring the well known fact that abundance measurements are dominated by systematics (appoximations on physics, uncertain atomic data, undetected or mis-estimated blends, poor continuum placement in crowded regions). We thus quote for all these species the \fei~ line-to-line scatter as representative uncertainty on [X/Fe], and then sum it in quadrature with either the \fei~ or the \feii~ line-to-line scatter for [X/Fe].

\end{itemize}

We do not present here an investigation of the dependence of the abundances from the chosen atmosphere parameters, however, as can be seen in \citet{francois07}, parameter sensitivity for stars of similar characteristics (\object{BS 17569-049}, Tab. 2 in \citealt{francois07}), is small compared to the uncertainties we quote, and can thus be neglected.

\end{appendix}
\end{document}